\documentclass[aps, prd, nofootinbib, preprintnumbers, showpacs, showkeys,preprint]{revtex4}
\usepackage{amsmath}
\usepackage{amssymb}
\allowdisplaybreaks

\usepackage{makeidx}
\usepackage{amsfonts}
\usepackage[usenames,dvipsnames]{pstricks}
\usepackage{subfigure}
\usepackage{epsfig}
\usepackage{pst-grad} 
\usepackage{mathtools}
\usepackage[colorlinks,hyperindex]{hyperref}
\usepackage{array}
\usepackage{tikz}
\usepackage{ulem}
\usetikzlibrary{decorations.markings,math,shapes.misc,patterns,arrows.meta,calc,patterns,angles,quotes}

\hypersetup
{	colorlinks,%
	citecolor=black,%
	linkcolor=black,%
	urlcolor=black,%
}

\usepackage{allpurpose}

\newcommand {\nn}{\nonumber}

\begin{document}

\title{Constraining the spacetime spin using time delay in stationary axisymmetric spacetimes}

\author{Haotian Liu}
\affiliation{School of Physics and Technology, Wuhan University, Wuhan, 430072, China}

\author{Junji Jia}
\email{junjijia@whu.edu.cn}
\affiliation{Center for Astrophysics \& MOE Key Laboratory of Artificial Micro- and Nano-structures, School of Physics and Technology, Wuhan University, Wuhan, 430072, China}


\begin{abstract}

Total travel time $t$ and time delay $\Delta t$ between images of gravitational lensing (GL) in the equatorial plane of stationary axisymmetric (SAS) spacetimes for null and timelike signals with arbitrary velocity are studied. Using a perturbative method in the weak field limit, $t$ in general SAS spacetimes is expressed as a quasi-series of the impact parameter $b$ with coefficients involving the source-lens distance $r_s$ and lens-detector distances$r_d$, signal velocity $v$, and asymptotic expansion coefficients of the metric functions. The time delay $\Delta t$ to the leading order(s) were shown to be determined by the spacetime mass $M$, spin angular momentum $a$ and post-Newtonian parameter $\gamma$, and kinematic variables $r_s,~r_d,~v$ and source angular position $\beta$.  When $\beta\ll \sqrt{aM}/r_{s,d}$, $\Delta t$ is dominated by the contribution linear to spin $a$. Modeling the Sgr A* supermassive black hole as a Kerr-Newman black hole, we show that as long as $\beta\lesssim 1.5\times 10^{-5}$ [$^{\prime\prime}$], then $\Delta t$ will be able to reach the $\mathcal{O}(1)$ second level, which is well within the time resolution of current GRB, gravitational wave and neutrino observatories. Therefore measuring $\Delta t$ in GL of these signals will allow us to constrain the spin of the Sgr A*.

\end{abstract}


\maketitle

\section{Introducing}

Nowadays time delay between gravitational lensing (GL) images has become a useful tool in astrophysics and cosmology. Time delay in GL of compact objects can be used to constrain their properties including mass, charge and distance to earth \cite{Virbhadra:2007kw,Bozza:2003cp,Zhao:2016kft,Wang:2019cuf}, distinguish black hole (BH) and naked singularity \cite{Virbhadra:2007kw,Sahu:2013uya}, and test theories of gravity \cite{Lu:2016gsf,Zhao:2017cwk,Zhao:2017jmv,Cao:2018lrd,Liu:2019pov,Lu:2019ush,Lu:2020kpo}. For GLs by galaxies or galaxy clusters, time delay can determine the Hubble parameter, matter density, dark matter substructure and dark universe parameters \cite{Oguri:2006qp,Keeton:2008gq,Coe:2009wt,Linder:2011dr,Mohammed:2014eca,Treu:2016ljm,Liao:2018ofi}.

The observed GL events are usually from light signals. However, with the observation of supernova neutrinos \cite{Hirata:1987hu, Bionta:1987qt, IceCube:2018dnn, IceCube:2018cha} and gravitational waves (GWs) \cite{Abbott:2016blz, Abbott:2016nmj, Abbott:2017oio, TheLIGOScientific:2017qsa, Monitor:2017mdv}, the astronomical observation entered the multimessenger era. Consequently, the time delays of neutrino and GW signals can be viewed as important supplements to time delay of light signals. Compared with time delay of light signals alone, the difference between time delays of light and neutrinos or light and GW signals can provide stronger constraints on the cosmology parameters \cite{Liao:2017ioi, Wei:2017emo, Fan:2016swi}. In addition, the time delay of these signals can determine the properties of test particles like mass ordering of neutrinos and velocity of GW \cite{Fan:2016swi, Jia:2017oar, Jia:2019hih}. Although it is known that neutrinos as well as GWs in some gravitational theories beyond General Relativity have non-zero masses, most of the previous works on their time delay treated them as null signals \cite{Fan:2016swi, Mena:2006ym, Eiroa:2008ks, Takahashi:2016jom}. It is obvious that time delay applicable to timelike signals will provide higher accuracy and therefore stronger constraints to spacetime and signal particle parameters when GWs and neutrinos are used as messengers \cite{Jia:2015zon,Pang:2018jpm}.

Previously, we showed that the time delay of timelike signals in spherically symmetric (SSS) spacetimes in the weak field limit is related to the asymptotic expansion coefficients of the metric functions, including spacetime mass $M$ and post-Newtonian parameter $\gamma$ etc \cite{Liu:2020td2}. However, the black hole (BH) no-hair conjecture implies that in general there exist another important parameter for BHs and potentially other compact objects, i.e., their spin angular momentum $a$. In this work, we will generalize our previous work in the SSS spacetimes to the time delay in the equatorial plane of arbitrary stationary axisymmetric (SAS) spacetimes. We will show that the time delay in the SAS case has a significant difference from that of the SSS spacetimes: the appearance of the spin dependant term at the very leading order. Applying the result to Kerr-Newman (KN) spacetime, we will show that the time delay between GL images due to the Sgr A* supermassive BH (SMBH) can be used to constrain the SMBH spin. Some other works calculated the Shapiro time delay (or the total travel time) in specific spacetimes for light signals. Ref. \cite{Sereno:2003nd}, \cite{Keeton:2005jd}, \cite{Wang:2014yya} and \cite{He:2016xiu,He:2016cya} calculated the Shapiro time delay in the Reissner-Nordstrom (RN), Schwarzschild, Kerr and KN spacetimes respectively for light signal. The time delay in the strong field limit was studied in Ref. \cite{Bozza:2003cp,Zhao:2016kft,Wang:2019cuf,Lu:2016gsf,Zhao:2017cwk,Zhao:2017jmv,Lu:2019ush}.

The paper is organized as follows. In Sec. \ref{sec:tt}, we use the perturbative method to obtain the total travel time $t$ in a quasi-series form of the impact parameter for signals with arbitrary velocities in general asymptotically flat SAS spacetimes. In Sec. \ref{sec:td}, time delay $\Delta t$ between two GL images is obtained to the leading order(s) using deflection angle that is accurate to the given order. In Sec. \ref{sec:tdkn}, we apply our results in general SAS spacetimes to the KN spacetime. We then model the Sgr A* SMBH as a KN BH and show how its spin $a$ can be determined using time delay between different GL images.

\section{Total travel time in SAS spacetimes} \label{sec:tt}

In this section we compute the total travel time in the equatorial plane of general SAS spacetimes using a perturbative method, which is essentially a combination and extension of those used in Refs. \cite{Huang:2020trl,Liu:2020td2} (see also \cite{Duan:2020tsq}). Therefore, we first recap some of the key steps of the method developed in these works and then calculate in details the total time in general SAS spacetimes.

We begin with the most general SAS metric, which can be described as
\begin{equation}\label{eq:sasmet}
  \dd s^2 = -A\dd t^2 + B \dd t \dd \varphi + C \dd \varphi^2 + D \dd r^2 + F \dd \theta^2,
\end{equation}
where $(t,~r,~\theta,~\varphi)$ are the coordinates and $A,~B,~C,~D,~F$ are metric functions depending only on $r$ and $\theta$. We choose the spherical coordinates $(r,~\theta)$ here rather than the cylindrical ones $(\rho,~z)$ since they allow us  to reduce to SSS spacetimes by simply setting $B=0$  \cite{Huang:2020trl, Sloane:1978}. We assume that the spacetime \eqref{eq:sasmet} permits motion of particles in a plane with fixed $\theta$, which can always be shifted to $\theta=\pi/2$ and called the equatorial plane. We then concentrate on motions in this plane, whose metric after suppressing the $\theta$ coordinate is
\begin{equation}\label{eq:sasmeteqc}
  \dd s^2 = -A(r) \dd t^2 + B(r) \dd t \dd \varphi + C(r) \dd \varphi^2 + D(r) \dd r^2.
\end{equation}

Using this metric, one can routinely obtain the geodesic equations,
\begin{align}
\dot{t}&= \frac{2 (L B+2 E C)}{B^2+4 A C}, \label{eq:tdeq}\\
\dot{\varphi}&=\frac{2 (2 L A-E B)}{B^2+4 A C}, \label{eq:phideq}\\
\dot{r}^2 &=\frac{\lb E^2 - \kappa A\rb \lb B^2+4AC\rb - \lb 2LA-EB\rb^2}{AD \left(B^2+4 A C\right)}, \label{eq:rdeq}
\end{align}
where $\kappa=1,~0$ respectively for timelike and null rays, and $\dot{~}$ stands for the derivative with respect to the proper time or affine parameter. $L$ and $E$ are two constants of motion due to the independence of the metric functions on $\varphi$ and $t$ respectively.
In asymptotically flat spacetimes, $L$ and $E$ can be interpreted respectively as the angular momentum and the energy of the massless particle or the unit mass of a massive particles. They can be further correlated to the impact parameter $b$ of the trajectory and asymptotic velocity $v$ of the massive particle,
\begin{equation}\label{eq:lerela}
L=(\textbf{p}\times \textbf{r})\cdot \hat{\textbf{z}}=\frac{bv}{\sqrt{1-v^2}},~E=\frac{1}{\sqrt{1-v^2}}.
\end{equation}
In the massless limit, note that although $L$ and $E$ diverge, $L/E=bv$ still holds. Throughout the paper, we allow $L$ and $b$ to carry signs: when the initial asymptotic approach of the signal is anticlockwise (or clockwise) with respect to the lens center, $L$ and $b$ are positive (or negative).

Using Eqs. \eqref{eq:tdeq} and \eqref{eq:rdeq}, then the total travel time for a signal from source at radius $r_s$ to detector at radius $r_d$ is
\begin{align} \label{eq:ttorif}
t=&\lsb \int_{r_0}^{r_s}+\int_{r_0}^{r_d}\rsb \frac{\sqrt{AD}\lb 2LB+4EC\rb}{\sqrt{B^2+4 A C }} \frac{\dd r}{\sqrt{\lb E^2 - \kappa A \rb \lb B^2+4AC\rb - \lb 2LA-EB\rb^2}} ,
\end{align}
where $A,~B,~C,~D$ here are functions of $r$, and $r_0$ is the closest approach of the trajectory. Furthermore, setting $\dot{r}|_{r=r_0}=0$ in Eq. \eqref{eq:rdeq}, the angular momentum $L$ can also be solved as a function of $r_0$
\begin{equation} \label{eq:angmemg}
L=\frac{E B(r_0) +s \sqrt{\lsb E^2 - \kappa A(r_0) \rsb \lsb B^2(r_0)+4A(r_0)C(r_0)\rsb} }{2 A(r_0)}
\end{equation}
where $s=+1,~-1$ respectively for prograde and retrograde motions of the signal. Note that in the relativistic limit, $v\to c$ and $E\to\infty$, and therefore $s=\mbox{sign}(L)=\mbox{sign}(b)$.
In application to GL observation, the impact parameter $b$ is often preferred over the closest approach $r_0$. Using Eqs. \eqref{eq:lerela} and \eqref{eq:angmemg}, we can establish a relation between them, as
\bea
\frac{1}{b}&=&\frac{2 A(r_0)\sqrt{E^2-\kappa}-B(r_0)E/b}{s\sqrt{\lsb 4A(r_0)C(r_0)+B(r_0)^2\rsb\lsb E^2-\kappa A(r_0)\rsb}} \label{eq:oobeq} \\
&\equiv&  p\lb b, \frac{1}{r_0}\rb. \label{eq:pdef}
\eea
Here in the last step we defined the right-hand side of Eq. \eqref{eq:oobeq} as a function $p$ of both $b$ and $1/r_0$. We can formally obtain $p(b,x)$'s inverse function $q(b,x)$ with respect to its second argument such that
\be
\frac{1}{r_0}=q\lb b ,~\frac{1}{b}\rb. \ee

Here we give $p(b,x)$ and $q(b,y)$ in Schwarzschild spacetime as an example. In Schwarzschild spacetime, metric function $B(r)=0$, so there is no dependence of $p(b,x)$ and $q(b,y)$ on their first parameter $b$. Using Eq. \refer{eq:pdef}, we have
\be
p_{Sch}\lb x \rb = \frac{x \sqrt{E^2-\kappa} }{\sqrt{\frac{E^2}{1-2Mx}-\kappa } }, \nn
\ee
and its corresponding inverse function $q(x)$ is
\be
q_{Sch}(x) = \frac{1}{6M}+\frac{\lb \mathrm{i}\sqrt{3} - 1 \rb\lsb E^2-\lb 1+12M^2x^2 \rb \kappa \rsb}{12M q_{f}(x) }-\frac{\lb 1+ \mathrm{i}\sqrt{3} \rb q_{f}(x) }{12M\lb E^2-\kappa \rb}, \nn
\ee
where the $\mathrm{i}$ is imaginary unit and
\begin{align}
q_{f}(x)=& \lcb E^6\lb 1-54M^2 x^2 \rb + 3 E^4 \kappa \lb 48M^2x^2 -1 \rb + 3 E^2 \kappa \lb 1-42M^2x^2\rb \right. \nonumber \\
  & +\kappa \lb 36M^2x^2-1\rb + 6\sqrt{3} \lcb M^2x^2\lb E^2-\kappa\rb^3 \lsb E^6\lb 27M^2x^2-1 \rb \right.\right. \nonumber \\
  &\left. \left. \left.  +3 E^4\kappa \lb 1-21M^2x^2\rb + E^2 \kappa \lb 44M^2x^2-3\rb + \kappa \lb 1-4M^2x^2 \rb^2 \rsb \rcb^{1/2} \rcb^{1/3}. \nn
\end{align}
Later on, substituting these functions $p(x)$ and $q(x)$ into Eqs. \eqref{eq:ttwithy} and \eqref{eq:yform} yields a total time $t$ that can be series expanded in powers of $1/b^n$ and so that the rest of the integral can be carried out.
For more complex spacetimes, the function $p(b,x)$ are always easy to obtain once the metric is given. We emphasis that it is usually much harder to find the analytical form of its inverse function $q(b,x)$, and more importantly, it is not necessary. What is needed in later calculation is only the series expansion of $p(b,x)$, which is always obtainable through the Lagrange inversion theorem from the series form of $p(b,x)$.

To carry out the integration in \eqref{eq:ttorif}, as in Ref. \cite{Huang:2020trl}, we then do a key change of variable from $r$ to $u$ which are linked by the relation
\be
\frac{1}{r}=q\lb b ,~\frac{u}{b}\rb.\label{eq:qinu} \ee
After some simple but slightly tedious algebra, the various terms in Eq. \eqref{eq:ttorif} then becomes \cite{Huang:2020trl,Liu:2020td2}
\begin{align}
&r_0\to 1,~ r_{s,d}\to b\cdot p\lb b,\ \frac{1}{r_{s,d}}\rb\equiv \sin \theta_{s,d},\label{eq:bptotheta} \\
&\frac{\sqrt{A(r)D(r)}}{\sqrt{B^2(r)+4A(r)C(r)}}\to\frac{\sqrt{A(1/q) D(1/q) }}{\sqrt{B^2(1/q) +4A(1/q) C(1/q) }},\label{eq:factorone}\\
&\frac{2LB(r)+4EC(r)}{2LA(r)-EB(r)}\to \frac{2bvB(1/q)+4C(1/q)}{2bvA(1/q)-B(1/q)},\label{eq:factortwo}\\
&\frac{2LA(r)-EB(r)}{\sqrt{\lsb E^2-\kappa A(r)\rsb\lsb B^2(r)+4A(r)C(r)\rsb -\lsb 2LA(r)-EB(r)\rsb^2}}\to\frac{su}{\sqrt{1-u^2}},\label{eq:factorthr} \\
&\dd r\to-\frac{1}{p_2(b,q)q^2}\frac{1}{b}\dd u, \label{eq:factorfou}
\end{align}
where the $\theta_{s,d}$ defined in Eq. \eqref{eq:bptotheta}, i.e.,
\be
\theta_{s,d}=\arcsin\lsb b\cdot p\lb b,\frac{1}{r_{s,d}}\rb\rsb \label{eq:thetasddef}
\ee
are indeed the apparent angles of the signal at the source and detector respectively \cite{Huang:2020trl}. $p_2(b,q)$ is the derivative of the function $p(b,q)$ with respect to it second argument $q$, which is given in Eq. \eqref{eq:qinu}.
Collecting these terms together, the total travel time \eqref{eq:ttorif} becomes
\begin{equation}\label{eq:ttwithy}
t=\lsb \int_{\sin\theta_s}^{1}+\int_{\sin\theta_d}^{1} \rsb y\lb b,\frac{u}{b}\rb \frac{\dd u}{u\sqrt{1-u^2}},
\end{equation}
where
\begin{align}  \label{eq:yform}
y\lb b,\frac{u}{b} \rb =& \frac{\sqrt{A(1/q) D(1/q) }}{\sqrt{B^2(1/q) +4A(1/q) C(1/q) }}
\frac{\lsb 2bvB(1/q)+4C(1/q)\rsb sb}{2bvA(1/q)-B(1/q)}\frac{1}{p_2(b,q)q^2}\lb\frac{u}{b}\rb^2 .
\end{align}
This $y\lb b,~u/b\rb$ depends on $u$ only through the ratio $u/b$. Thus, we can expand it with respect to its second argument to find a series form
\begin{equation}\label{eq:bserub}
y\lb b,\frac{u}{b}\rb=\sum_{n=n_0}^{\infty} y_n(b)\frac{u^n}{b^n},
\end{equation}
where $y_n(b)$ are the expansion coefficients. The leading index of this series, $n_0$, is determined by the asymptotic behavior of the metric functions in Eq. \eqref{eq:sasmeteqc}. If they satisfy the asymptotic expansion \eqref{eq:metasyf}, then one can show that $n_0=-1$.
Finally, changing the integration variable in Eq. \eqref{eq:ttwithy} from $u$ to $\xi$ by $u=\sin \xi$ , the total time delay \eqref{eq:ttwithy} becomes
\begin{equation}\label{eq:ttynutoxi}
t=\lsb \int_{\theta_s}^{\frac{\pi}{2}}+\int_{\theta_d}^{\frac{\pi}{2}} \rsb \sum_{n=n_0}^{\infty} y_n\lb b\rb \frac{\sin^{n-1} \xi}{b^n} \dd \xi.
\end{equation}
The integrability of this time delay now is clear because the functions $\sin^{n-1}\xi$ can always be integrated to find
\begin{align}\label{eq:lnform}
l_n(\theta_s,~\theta_d) \equiv& \lsb \int_{\theta_s}^{\frac{\pi}{2}} +\int_{\theta_d}^{\frac{\pi}{2}} \rsb\sin^{n-1} \xi \dd \xi  \\
 =& \begin{cases}
\displaystyle \sum_{i=s,d} \cot\theta_i,&n=-1,\\
\displaystyle \sum_{i=s,d} \ln\lsb\cot\lb\frac{\theta_i}{2}\rb\rsb,&n=0,\\
\displaystyle \sum_{i=s,d} \frac{(n-2)!!}{(n-1)!!} \left(\frac{\pi}{2}-\theta_i
    +\cos\theta_i\sum_{j=1}^{[(n-1)/2]} \frac{(2j-2)!!} {(2j-1)!!}\sin^{2j-1} \theta_i\right),&n=1,3,\cdots,\\
\displaystyle \sum_{i=s,d} \frac{(n-2)!!}{(n-1)!!} \cos\theta_i \left(1
    +\sum_{j=1}^{[(n-1)/2]} \frac{(2j-1)!!}{(2j)!!} \sin^{2j}\theta_i\right),&n=2,4,\cdots.
\end{cases} \nonumber
\end{align}
Substituting these into Eq. \eqref{eq:ttynutoxi}, we finally find the formal total travel time in a perturbative form
\begin{equation}\label{eq:ttynln}
t=\sum_{n=-1}^{\infty} y_n\lb b\rb \frac{l_n}{b^n}.
\end{equation}

A few comments regarding the above procedure and the result \eqref{eq:ttynln} are now in order. Firstly, in expanding $y(b,u/b)$ to series \eqref{eq:bserub}, the explicit form of the inverse function $q(b,u/b)$ of $p(b,x)$ is seemingly needed, while it is not always possible to do so for some even simple functions. However indeed here the true $q(b,u/b)$ is not really necessary because what we need is only its expansion and this can be achieved through the Lagrange inverse theorem using $p(b,x)$ alone. Secondly, the expansion \eqref{eq:bserub} is actually carried out in the small $u$ limit as a Laurent series. It can also be viewed as a quasi-series of large $b$, for which the range of convergence can be shown to be $(b_c,~\infty)$. Here $b_c$ is the critical impact parameter below which the particle will not escape to infinity. Mathematically, it is the largest singular $b$ of the function $y(b,u/b)$. Thirdly, as in the case of SSS spacetime in Ref. \cite{Liu:2020td2}, here for any value of the impact parameter inside the range of convergence, the perturbative result \eqref{eq:ttynln} will also be able to reach any desired accuracy if the series is truncated at high enough order. Furthermore, in the large $b$ limit, the coefficients of the series \eqref{eq:ttynln}, $y_n(b)$, is completely determined by the behavior of the metric functions in the asymptotic region, as was shown in Ref. \cite{Huang:2020trl,Liu:2020td2}.

Next we will compute the first few coefficients $y_n(b)$ for asymptotically flat spacetimes and then the total time \eqref{eq:ttynln} in the weak field limit. Metric functions of the asymptotically flat SAS spacetimes always have an asymptotic expansion of the form
\begin{equation}\label{eq:metasyf}
A(r)\to 1+\sum_{i=1}\frac{a_i}{r^i},~rB(r)\to \sum_{i=0}\frac{b_i}{r^i},~\frac{C(r)}{r^2}\to 1+\sum_{i=1}\frac{c_i}{r^i},~D(r)\to 1+\sum_{i=1}\frac{d_i}{r^i},
\end{equation}
where $a_i,~b_i,~c_i,~d_i$ are constants. Without losing generality, here $a_1$ can be identified with the minus 2 times the ADM mass $M$ of the spacetime.
Substituting \eqref{eq:metasyf} into $y(b,u/b)$ in Eq. \eqref{eq:bserub},  $y_n(b)$ for $n=-1,0,1,2$ are easily found to be
\begin{align}  \label{eq:sasynexp}
y_{-1}(b)=&\frac{s}{v},  \\
y_0(b)=&\frac{a_1(1-2 v^2)+d_1 v^2}{2 v^3}, \\
y_1(b)=&s\lsb\frac{8 a_1^2-4 a_1 (c_1+d_1)-8 a_2-(c_1-d_1)^2+4 c_2+4 d_2}{8 v}\right.\nonumber\\
&\left. +\frac{2 b_0 v (-4 a_1+c_1+d_1)+4 b_1 v}{8 v^3} \frac{1}{b} + \frac{b_0^2}{4 v^3}\frac{1}{b^2}\rsb,\\
y_2(b)=&\frac{b_0b}{2} +\frac{1}{16 v^7} \lcb -a_1^3 \left(16 v^6+48 v^4+18 v^2+1\right)+a_1^2 v^2 \lsb 2 c_1 \left(8 v^4+32 v^2+11\right) \right. \right. \nonumber \\
   & \left. + d_1 \left(8 v^4+8 v^2-1\right)\rsb +a_1 \lsb 4 a_2 v^2 \left(8 v^4+20 v^2+5\right)-v^4 \left(24 c_1^2+4 c_1 \left(2 d_1 v^2+d_1\right) \right. \right. \nonumber \\
   &\left. \left. + 8 c_2 \left(2 v^2+7\right)-\left(2 v^2+1\right) \left(d_1^2-4 d_2\right)\right)\rsb +v^4 \lsb -4 a_2 \left(4 c_1 v^2+14 c_1+2 d_1 v^2+d_1\right) \right. \nonumber \\
   & \left. \left. - 16 a_3 \left(v^2+2\right)+v^2 \left(24 c_1 c_2-2 c_1 d_1^2+8 c_1 d_2+8 c_2 d_1+40 c_3+d_1^3-4 d_1 d_2+8 d_3\right)\rsb \rcb \nonumber \\
   & + \frac{1}{8 b v^6} \lcb a_1^2 b_0 \left(36 v^4+60 v^2+11\right)-2 a_1 v^2 \lsb b_0 \left(c_1 \left(20 v^2+26\right)+4 d_1 v^2+d_1\right) \right. \right. \nonumber \\
   & \left. +14 b_1 \left(v^2+1\right)\rsb -28 a_2 b_0 \left(v^2+1\right) v^2+v^4 \lsb b_0 \left(12 c_1^2+4 c_1 d_1+32 c_2-d_1^2+4 d_2\right) \right. \nonumber \\
& \left. \left. +4 b_1 (8 c_1+d_1)+20 b_2\rsb \rcb +\frac{3 b_0 \lsb v^2 \lcb b_0 (12 c_1+d_1)+14 b_1\rcb -a_1 b_0 \left(16 v^2+11\right)\rsb}{8 b^2 v^5}+\frac{19 b_0^3}{8 b^3 v^4}.
\end{align}
Here $y_{-1}$ and $y_0$ agree with the corresponding SSS spacetime result in Ref. \cite{Liu:2020td2} and the parameters containing the spin of the spacetime, including $b_{0,1,2},~c_{1,2,3}$ and $d_{2,3}$, only appear in $y_1,~y_2$ and orders above.
Higher order results can also be found without difficulty but are not needed in the following calculations and therefore will not be shown. In general, one can show that these coefficients always take a form
\begin{align}
y_n(b)=& M^{n+1} \sum_{j=-1}^{n+1} y_{n,j} \frac{M^j}{b^j},~n\geqslant 2,
\end{align}
where $y_{n,j}$ are polynomials of dimensionless quantities $a_i/M^i,~ b_i/M^{i+2},~c_i/M^i,~d_i/M^i$.

In next section, we will use this total time to compute the time delay between different images of the GL in the weak field limit. In this limit, we have $r_{s,d}\gg b\gg M$ and therefore there will be two small parameters $b/r_{s,d}$ and $M/b$. Now the factors $y_n(b)$ and $1/b^n$ in series \eqref{eq:ttynln} are already in power forms of $M/b$. One can straightforwardly expand the only other factors $l_n$ in Eq. \eqref{eq:ttynln} into a series of small $b/r_{s,d}$ and $M/b$ too. The result for the first four $l_n~(n=-1,0,1,2)$ are
\begin{align} \label{eq:saslnexp}
l_{-1}=&\sum_{i=s,d}\lsb s\lb \frac{r_i}{b}-\frac{b}{2 r_i} \rb + \frac{s}{b} \frac{c_1 v^2-a_1}{2 v^2} +\frac{1}{b^2}\frac{sb_0}{2v}+ \mathcal{O}\lb \frac{b^3}{r_i^3}, ~\frac{Mb}{r_i^2} , ~\frac{M^2}{br_i} \rb \rsb, \\
l_0=&\sum_{i=s,d} \lsb \ln \lb \frac{2r_i}{b} \rb + \frac{c_1 v^2-a_1}{2 v^2} \frac{1}{b} \frac{b}{r_i} + \mathcal{O}\lb \frac{b^2}{r_i^2},~\frac{M^2}{b r_i}  \rb \rsb, \\
l_1=&\sum_{i=s,d} \lsb\frac{\pi}{2}-  \frac{sb}{r_i}  + \mathcal{O}\lb \frac{b^3}{r_i^3},~ \frac{Mb}{r_i^2} \rb \rsb, \\
l_2=& \sum_{i=s,d}\lsb 1 + \mathcal{O}\lb \frac{b^2}{r_i^2} \rb \rsb. \end{align}
Again, expansion of high order $l_n$ are also easy but not necessary in the following computations.

Substituting Eqs. \eqref{eq:sasynexp} and \eqref{eq:saslnexp} into Eq. \eqref{eq:ttynln}, the total travel time expressed completely as series of $M/b$ and $b/r_{s,d}$ becomes
\begin{align} \label{eq:ttgenfth}
t=&\sum_{i=s,d}\lcb \frac{b}{v}\lsb \frac{r_i}{b}-\frac{b}{2r_i}+ \frac{1}{b} \frac{c_1v^2-a_1}{2 v^2}\rsb  + \frac{\lb d_1-2 a_1 \rb v^2+a_1}{2 v^3} \lsb \ln \left(\frac{2 r_i}{sb}\right) +\frac{c_1 v^2-a_1}{2 v^2}\frac{1}{r_i}\rsb\right. \nonumber \\
 & + \frac{8 a_1^2 - 4 a_1 (c_1+d_1)-8 a_2-(c_1-d_1)^2+4 c_2+4 d_2}{8 v}  \frac{s\pi}{2}  \frac{1}{b} + \frac{b_0}{2 b}\lb 1+\frac{1}{v^2}\rb \nonumber \\
 &\left. + \mathcal{O}\lb \frac{M^3}{b^2},\frac{b^4}{r_i^3}\rb\rcb .
\end{align}
Note that the $i$-th $(i=1,2,3,4)$ term in Eq. \eqref{eq:ttgenfth} originates from the $i$-th term of Eq. \eqref{eq:ttynln} and we have ignored the order higher than $\mathcal{O}\lb M^3/b^2, b^4/r_i^3\rb$. Clearly, this total travel time is in a full series form of $M/b$ and $b/r_i$. If higher accuracy than Eq. \eqref{eq:ttgenfth} is needed, then one only needs to include $y_n,~l_n$ to higher orders and keep more terms in their expansion. Moreover, going to higher order will also allow one to study the effect of higher order PPN parameters on the total time and time delay. Note if we substitute the velocity $v$ to 1 and the general SSS, RN, Kerr and KN metrics into this total time, then Eq. (107) of Ref. \cite{Keeton:2005jd}, Eq. (12) of Ref. \cite{Sereno:2003nd}, Eq. (22) (after expansion in the large $r/r_0$ limit) of Ref. \cite{Wang:2014yya}, Eq. (20) of Ref. \cite{He:2016xiu} and (27) (after expansion in the large $X_{A,B}/b$ limit) of Ref. \cite{He:2016cya} will be produced respectively.

\section{Time delay in general SAS spacetimes} \label{sec:td}

\begin{figure}[htp]
\includegraphics[width=0.6\textwidth]{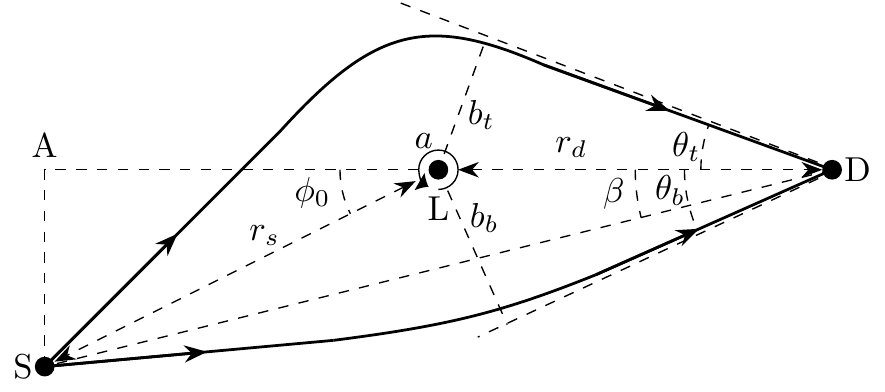}
\caption{The GL in an SAS spacetimes. S, L, D are the source, lens and detector respectively. $b_t$ and $b_b$ are the impact parameters for the top and bottom paths. $beta$ is the source's angular position assuming that the gravitational center was not present. The spin angular momentum of the lens in this illustration is anticlockwise. We choose the sign $\epsilon=\mbox{sign}(\phi_0)=\mbox{sign}(\beta)=+1$ (and $-1$) when $\phi_0$ and $\beta$ are counterclockwise (and clockwise) against the observer-lens axis.}  \label{fig:lensing}
\end{figure}

For gravitationally lensed source in the equatorial plane of the SAS spacetime, there will always be one primary image (comparing to the multiple relativistic images in the strong field limit \cite{Virbhadra:2008ws}) on each side of the lens (see Fig. \ref{fig:lensing}). The apparent angles $\theta_b$ and $\theta_t$ of the images and the time delay between the images are of interest to astronomy observations.
Using lens equations that are accurate for the SAS spacetimes, in this section we will solve the image apparent angles and their corresponding impact parameters, from which we can further solve the time delay. The key result we will show is that the time delay $\Delta t$ to the leading order(s) will only depend on three parameters of the metric expansion \eqref{eq:metasyf}, the spacetime mass $M$, the spacetime spin $a$ and the parameter $\gamma$ in the parameterized post-Newtonian (PPN) formalism of gravity.

We first establish the GL equations which link the source's angular position when the gravitational center was not present, i.e. $\beta$, and its apparent angles $\theta_t$ and $\theta_b$ from the top and bottom paths respectively (see Fig. \ref{fig:lensing}).
Previously this was usually done
using the deflection angle without the finite distance correction and in the small angle approximation, $\beta,~\phi_0\ll 1$ \cite{Bozza:2008ev}. However here we would like to use geometric equations and deflection angle that are as accurate as possible. To do this,
we first link $\beta$ to $\phi_0$ in Fig. \ref{fig:lensing}. Using the triangles  $\triangle$SLA and $\triangle$SDA, we have a relation
\begin{equation}\label{eq:angrela}
(r_d+r_s\cos \phi_0) \tan \beta = r_s \sin \phi_0 .
\end{equation}
The $\phi_0$ then can be connected to both the impact parameters $b_t$ from the top side and $b_b$ from the bottom side through the change of the angular coordinate $\Delta \varphi(b_{b,t})$ from the corresponding side
\begin{equation}\label{eq:changeq}
\pm \pi + \phi_0 = \pm \Delta \varphi (b_{b,t})
\end{equation}
where upper sign ``$+$'' (or lower sign ``$-$'') is for the subscript $b$ (or $t$). $\Delta \varphi (b_{b,t})$ to the leading non-trivial order in an SAS spacetimes were found in Ref. \cite{Huang:2020trl} (see also \cite{Jia:2020xbc}) as
\begin{equation}\label{eq:changmet}
\Delta \varphi (b_{b,t}) \approx \pi + \frac{d_1v^2-a_1}{ b_{b,t} v^2}-b_{b,t}\lb \frac{1}{r_s}+\frac{1}{r_d}\rb.
\end{equation}
This $\Delta \varphi(b_{b,t})$ takes into account the finite distance effect of the source and detector and therefore provides a natural way to involve the source and lens distances into the GL equation. Note that \eqref{eq:changmet} ignores the terms of order $\mathcal{O}(M^2/b^2,~b^2/r_{s,d}^2)$ or higher, whose contribution to the final time delay is negligible.

Solving Eqs. \eqref{eq:changeq}-\eqref{eq:changmet}, we can obtain the impact parameters $b_{b,t}$ and link them to $r_{s,d}$ and $\phi_0$
\begin{equation}\label{eq:impabts}
b_{b,t}=\sqrt{\frac{(d_1 v^2 - a_1) r_d r_s}{\eta v^2 (r_d + r_s)}} \lsb \epsilon\pm \sqrt{\eta+1} \rsb,
\end{equation}
where $\epsilon=\mbox{sign}(\phi_0)=\mbox{sign}(\beta)$ and
\begin{equation}\label{eq:etainb}
\eta = \frac{4 (d_1 v^2 - a_1)(r_d + r_s)}{\phi_0^2 v^2 r_d r_s}.
\end{equation}
Once $b_{b,t}$ is known, one can immediately obtain the apparent angle of the image though relation \eqref{eq:thetasddef} as
\be \label{eq:thetabt}
\theta_{b,t}=\sin^{-1}\lsb b_{b,t}\cdot p\lb b_{b,t}, \frac{1}{r_d}\rb\rsb.
\ee

Note that in the above analysis, the Eqs. \eqref{eq:angrela}-\eqref{eq:changmet} when combined plays the rule of conventional well-known lens equation in the weak field limit
\be
\beta=\frac{b}{r_d}-\frac{r_s}{r_s+r_d}\frac{4M}{b}, \label{eq:convleq}
\ee
where the approximation $b=\theta r_d$ and the deflection angle at first order $\Delta\phi=4M/b$ is used. Only after combining these equations together but not any of them alone, the impact parameters $b_{b,t}$ from two sides can be solved, as given in Eq. \eqref{eq:impabts}. Moreover, if one is interested in the apparent angle $\theta_{b,t}$, then $b_{b,t}$ will have to be further substituted into Eq. \eqref{eq:thetabt}. We have verified explicitly for the Sgr A* data used in Sec. \eqref{sec:tdkn} that Eqs. \eqref{eq:angrela}-\eqref{eq:changmet} and \eqref{eq:thetabt} will generate the same values of $b_{b,t}$ and $\theta_{b,t}$, if the we kept also first order deflection angle in Eq. \eqref{eq:changmet}, as the conventional lens equation \eqref{eq:convleq}.

For the time delay between the images from two sides, substituting $b_{b,t}$ into Eq. \eqref{eq:ttgenfth} and subtracting each other, we obtain
\begin{align}\label{eq:tdgenform}
\Delta t = & \frac{\sqrt{\eta +1}}{\eta}\cdot \frac{2 \epsilon\left(d_1 v^2-a_1\right)}{  v^3} + \frac{ \left( d_1 -2 a_1\right) v^2+a_1}{v^3} \ln \left(\frac{\sqrt{\eta +1}+\epsilon }{\sqrt{\eta +1}-\epsilon }\right)  \nonumber \\
   & + \frac{\epsilon\pi \sqrt{r_s+r_d} }{4\sqrt{\eta (d_1 v^2 - a_1 ) r_s r_d} }\lsb 8 a_1^2-4 a_1 (c_1+d_1)-8 a_2-c_1^2+2 c_1 d_1+4 c_2-d_1^2+4 d_2\rsb \nonumber \\
   & +\sqrt{\frac{\eta+1}{\eta}}\cdot\frac{2 b_0v \sqrt{(r_s+r_d)} }{\sqrt{(d_1 v^2 - a_1 ) r_s r_d} }\lb 1+\frac{1}{v^2}\rb + \mathcal{O}\lb \phi_0\sqrt{\frac{M^3}{r_{s,d}}},\phi_0^4r_{s,d} \rb.
\end{align}
The four terms in this equation are respectively from the first to fourth term of Eq. \eqref{eq:ttgenfth} and we only keep the results of order $\mathcal{O}\lsb (1+1/\eta)\sqrt{M^3/r_{s,d}}\rsb$ or lower.
Similar to the SSS case in Ref. \cite{Liu:2020td2}, the contributions from the first two terms in Eq. \eqref{eq:tdgenform} dominate the third term when $\eta\gg1$, i.e. $|\phi_0|\ll \sqrt{M/r_{s,d}}$. On the other hand, when $\eta\ll 1$, i.e. $|\phi_0|\gg \sqrt{M/r_{s,d}}$, the first term will be much larger than the second and third terms. Unlike the SSS spacetime, the fourth term of Eq. \eqref{eq:tdgenform} is new because of the appearance of the spin angular momentum $b_0$ in the SAS spacetime. A comparison of the fourth term with the first three terms immediately tells that when
\be \label{eq:fourlargecond}
\eta\gg \frac{r_{s,d}}{M}\cdot \frac{M^2}{b_0}, ~\text{i.e.}~ |\phi_0|\ll \frac{M}{r_{s,d}}\lb \frac{b_0}{M^2}\rb^{1/2},\ee
the former will be much larger than the latter.
Since the fourth term is linear in $b_0$, this implies that when condition \eqref{eq:fourlargecond} is satisfied, the time delay will critically depend on the spacetime spin.
In Sec. \ref{sec:tdkn}, we will use this time delay to constrain the spin of the Sgr A* SMBH.

\begin{figure}[htp]
\begin{center}
\includegraphics[width=0.45\textwidth]{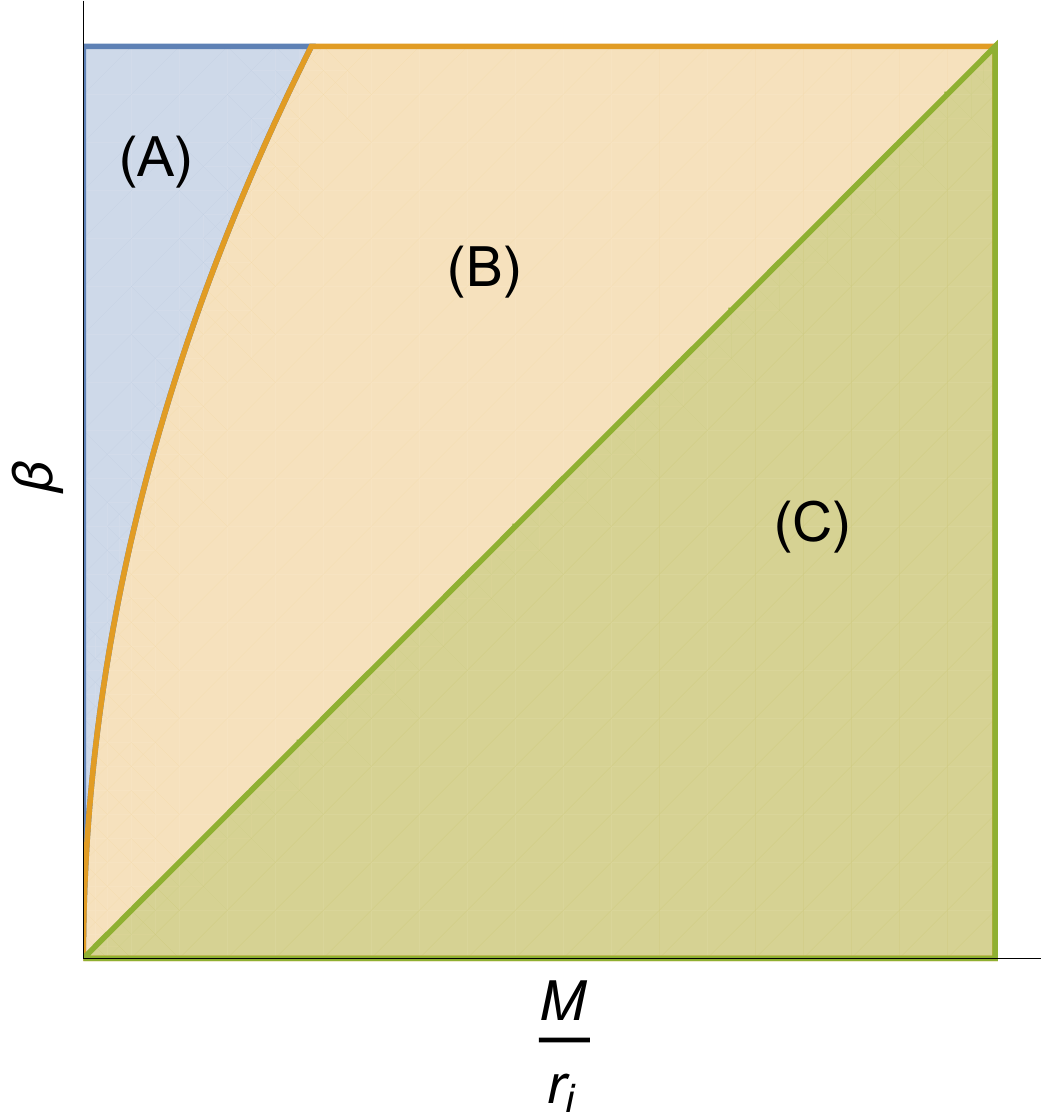}
\caption{The partition of the parameter space spanned by $M/r_{s,d}$ and $\beta$. The regions (A) and (B) are separated by curve $\beta\approx \sqrt{M/r_{s,d}}$, and (B) and (C) by $\beta\approx \sqrt{b_0}/r_{s,d}$. The relation $b_0/M^2\lesssim \mathcal{O}(1)$ is implicitly assumed. \label{fig:parapart}}
\end{center}
\end{figure}

Combining all these analysis, we see that in the entire parameter space spanned by $(M/r_{s,d},~\phi_0)$, the time delay can be well approximated by the first, second and fourth terms of Eq. \eqref{eq:tdgenform}. On the other hand, since the angle $\beta$ is more readily linked to GL observables than $\phi_0$, it is also desirable to obtain a time delay expressed in terms of $\beta$. To do this, we can directly use Eq. \eqref{eq:angrela} to replace $\phi_0$ by $\beta$ and obtain the time delay to the leading order(s)
\begin{align}\label{eq:tdgenformbeta}
\Delta t = & \frac{\sqrt{\eta(\beta,v)+1}}{\eta(\beta,v)} \cdot \frac{2\epsilon \left(d_1 v^2-a_1\right)}{ v^3} + \frac{ \left( d_1 -2 a_1\right) v^2+a_1}{v^3} \ln \left(\frac{\sqrt{\eta(\beta,v)+1}+\epsilon}{\sqrt{\eta(\beta,v)+1}-\epsilon }\right)  \nonumber \\
& +\sqrt{\frac{1+\eta(\beta,v)}{\eta(\beta,v)}}\cdot \frac{2 b_0v \sqrt{(r_s+r_d)} }{\sqrt{(d_1 v^2 - a_1 ) r_s r_d} } + \mathcal{O}\lb \beta\sqrt{\frac{M^3}{r_{s,d}}},\beta^4r_{s,d} \rb
\end{align}
where
\begin{equation}\label{eq:etainbbeta}
\eta(\beta,v)=\frac{4 (d_1 v^2 - a_1) r_s}{\beta^2 v^2 (r_s + r_d) r_d}.
\end{equation}
Similarly, the previous discussion about the dominance of each term(s) in Eq. \eqref{eq:tdgenform} according to the relation between $M/r_{s,d}$ and $\phi_0$ can also apply to the parameters $M/r_{s,d}$ and $\beta$ in Eq. \eqref{eq:tdgenformbeta}. In Fig. \ref{fig:parapart} we plot the partition of parameter space spanned by $(M/r_{s,d},~\beta)$ according to the relative size of each term of Eq. \eqref{eq:tdgenformbeta}. In region (A), the 3rd term > the 1st term > the 2nd term, while in region (B) the 3rd term > the 1st term $\approx$ the 2nd term, and in region (C) the 1st term $\approx$ the 2nd term > the 3rd term. Eq. \eqref{eq:tdgenformbeta} is one of the key result of this paper. Only three parameters from the spacetime metric expansion \eqref{eq:metasyf}, $a_1,~b_0$ and $d_1$, appear in it. As known to general SAS spacetimes, they are simply equivalent to the ADM mass $M$, spin angular momentum $a$ of the spacetime and the $\gamma$ parameter in the PPN formalism of gravity \cite{Bardeen:1973,Weinberg:1972kfs}
\be
a_1=-2M, ~b_0=-4aM,~d_1=2\gamma M.
\ee
Therefore we conclude that to the leading order(s), the time delay $\Delta t$ are determined by, in addition to kinematic variables $r_{s,d},~\beta,~v$, these three parameters of the spacetime.

The time delay \eqref{eq:tdgenformbeta} applies to signals of all velocity. For relativistic timelike signals, in order to see more clearly the effect of velocity, we can expand $\Delta t$ around the speed of light. The result to the first non-trivial order is
\begin{align} \label{eq:tdexpv}
\Delta t(v\to 1)=&\Delta t(v=1)\nn\\
&+\lcb \frac{2 \epsilon \lsb d_1 \lb \eta(\beta,1)+1 \rb - a_1 \lb 2 \eta(\beta,1)+1 \rb \rsb }{\eta(\beta,1) \sqrt{\eta(\beta,1)+1}}\right.\nn\\
& + \lsb (a_1+d_1) \ln \lb \frac{\sqrt{\eta(\beta,1)+1}+\epsilon}{\sqrt{\eta(\beta,1)+1}-\epsilon }\rb +\frac{2 \epsilon a_1 }{\sqrt{\eta(\beta,1)+1}}\rsb\nn\\
&\left. +\frac{4 b_0 \lsb a_1+d_1\lb \eta(\beta,1)+1 \rb\rsb  \sqrt{r_d+r_s} }{(d_1-a_1)^{3/2} \sqrt{ \eta(\beta,1) \lsb \eta(\beta,1)+1\rsb r_d r_s }}\rcb (1-v)\\
&\equiv \Delta t(v=1)+\Delta t_{c-v}.
\end{align}
Here we define $\Delta t_{c-v}$ as the deviation of the timelike signal's time delay from that of the null signal. $\Delta t_{c-v}$ has a particular advantage in GW/GRB dual lensing: because it is independent of the GW and GRB emission time difference and therefore can be used to constrain the GW speed \cite{Fan:2016swi,Jia:2019hih}.

\section{Time delay in KN spacetime and the spin of Sgr A* SMBH} \label{sec:tdkn}

We now apply the time delay  \eqref{eq:tdgenformbeta} to the KN spacetime and use it to constrain the spin of the Sgr A* SMBH.
We will first find out the time delay in KN spacetime and then substitute values of necessary parameters associated with Sgr A* to show how the time delay is related to various parameters, especially the spacetime spin angular momentum.

The metric of the KN spacetime is given by
\begin{align}
\dd s^2=&-\frac{\Delta-a^2\sin^2\theta}{\Sigma}\dd t^2+\frac{(a^2+r^2)^2-a^2\Delta \sin^2\theta}{\Sigma}\sin^2\theta\dd \phi^2-\frac{2a\sin^2\theta(a^2-\Delta+r^2)}{\Sigma}\dd t\dd \phi \nonumber \\
&+\frac{\Sigma}{\Delta}\dd r^2+\Sigma\dd \theta^2,
\end{align}
where
\be
\Sigma(r,\theta)=r^2+a^2\sin^2\theta,~\Delta(r,\theta)=r^2-2Mr+a^2+Q^2
\ee
and $M,~Q,~a=J/M$ are respectively the total mass, total charge and the specific spin angular momentum of the spacetime.
From this, one can immediately read off the metric functions in the equatorial plane and consequently find their asymptotic expansions in the form of Eq. \eqref{eq:metasyf} with coefficients
\begin{equation} \label{eq:kncoeff1}
\begin{split}
&a_1=-2M,~a_2=Q^2,\\
&b_0=-4aM,~b_1=2aQ^2,  \\
&c_2=a^2,~c_3=2a^2M, \\
&d_1=2M~(\text{i.e.}~\gamma=1),~d_2=4M^2-a^2-Q^2,~d_3=4M\lb 2 M^2 - a^2 - Q^2 \rb
\end{split}
\end{equation}
and all other coefficients equal to zero. Substituting these into Eq. \eqref{eq:tdgenformbeta}, the time delay in KN spacetime becomes
\begin{align} \label{eq:dtkn}
\Delta t_K = & \frac{\sqrt{\eta_K+1}}{\eta_K}\cdot \frac{4\epsilon M(1+v^2)}{ v^3}+\frac{2 M \left(3 v^2-1\right) }{v^3}\ln \lb \frac{\sqrt{\eta_K+1}+\epsilon }{\sqrt{\eta_K+1}-\epsilon }\rb  \nonumber \\
& - \sqrt{\frac{\eta_K+1}{\eta_K}}\cdot \frac{4 a  \sqrt{2(1+v^2)M( r_s+r_d)}}{v\sqrt{ r_s r_d} }  + \mathcal{O}\lb \beta\sqrt{\frac{M^3}{r_{s,d}}},\beta^4r_{s,d} \rb,
\end{align}
where
\be
\eta_K=\eta_K(\beta,v)= \frac{8M(1+ v^2) r_s}{\beta^2 v^2 (r_s + r_d) r_d}. \label{eq:knbeta}
\ee
Note that the charge $Q$ appears only in the third term in Eq. \eqref{eq:ttgenfth}, which is at least an order $\sqrt{M/r_{s,d}}$ smaller than the first and/or second terms and consequently is negligible in Eq. \eqref{eq:dtkn}.
Most importantly, it is seen that as pointed out in Sec. \ref{sec:td},  when $\beta\to0$, $\eta_K(\beta,v)\to\infty$ and the third term in Eq. \eqref{eq:dtkn} dominates
\begin{align} \label{eq:tdknps}
\Delta t_K(\beta\to 0)\approx -\frac{4 a \sqrt{2(1+v^2) M (r_d+r_s)}}{v\sqrt{ r_d r_s }} + \mathcal{O}\lb \frac{aM}{r_{s,d}}\rb,
\end{align}
which is linear to the spacetime spin $a$.

For long time, the spin of the Sgr A* has not been well constrained due to the relatively low accretion rate comparing to many other SMBHs \cite{Andreas:2018}.
In this work we model the Sgr A* SMBH as a KN BH with GL happening in its equatorial plane. The signal sources can be stars for light signal, binary mergers for GWs and supernovas for neutrinos. In particular, one can verify easily from Eq. \eqref{eq:dtkn} that when other parameters ($r_d,~v,~M,~a$) are fixed and $\beta$ is small, the smaller the $r_s$, the larger the time delay. Therefore it is advantageous to consider sources that are as close to the BH as possible and which still allow the weak field limit to hold. Such sources at least include some stars in the S star category near the Sgr A*, such as the star S39 whose orbit has an edge-on inclination of $89.36\pm0.73 ^\circ$ with respect to the plane of the sky \cite{2017ApJ...837...30G}. Using the orbit data of S39, we can find that its $r_s\approx  1.113\times 10^{-3}~\text{[pc]}\equiv r_{S39}$ when it is on far side of the SMBH.

\begin{figure}[htp]
\begin{center}
\includegraphics[width=0.45\textwidth]{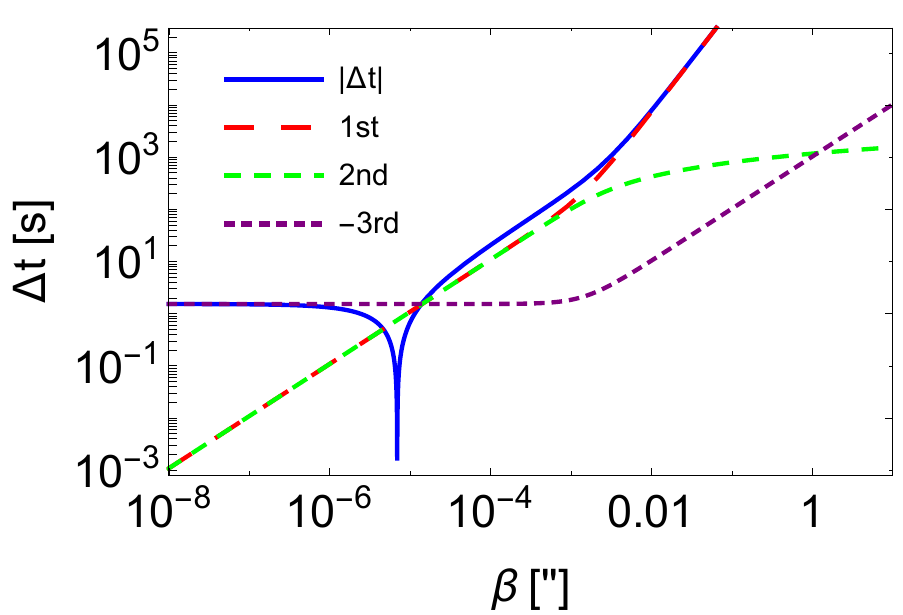}~~
\includegraphics[width=0.45\textwidth]{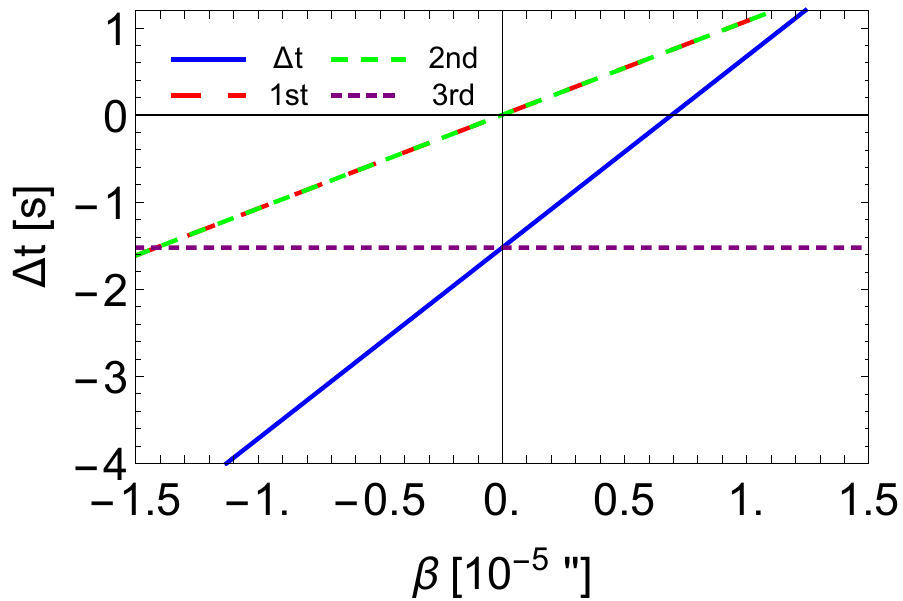}\\
~~~~~(a)\hspace{7cm}(b)\\
\includegraphics[width=0.45\textwidth]{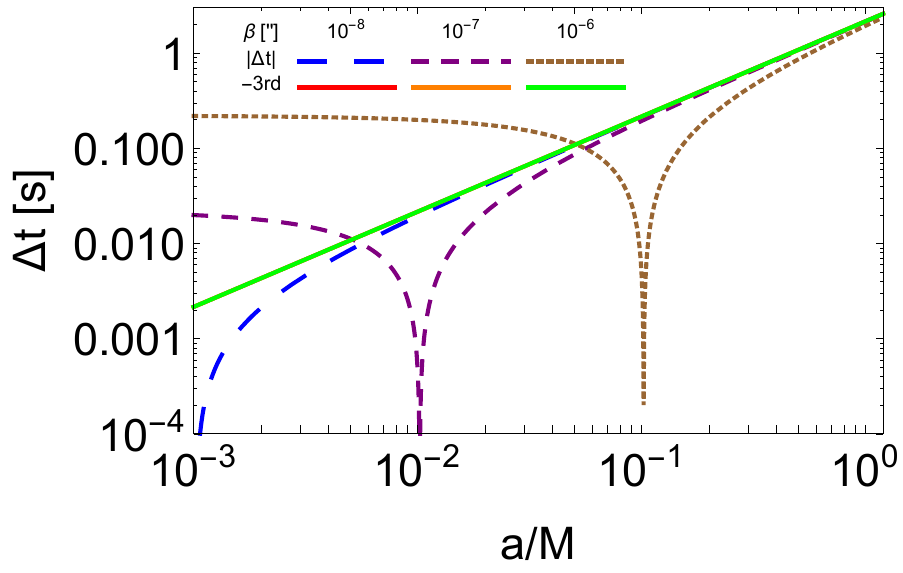}~~
\includegraphics[width=0.42\textwidth]{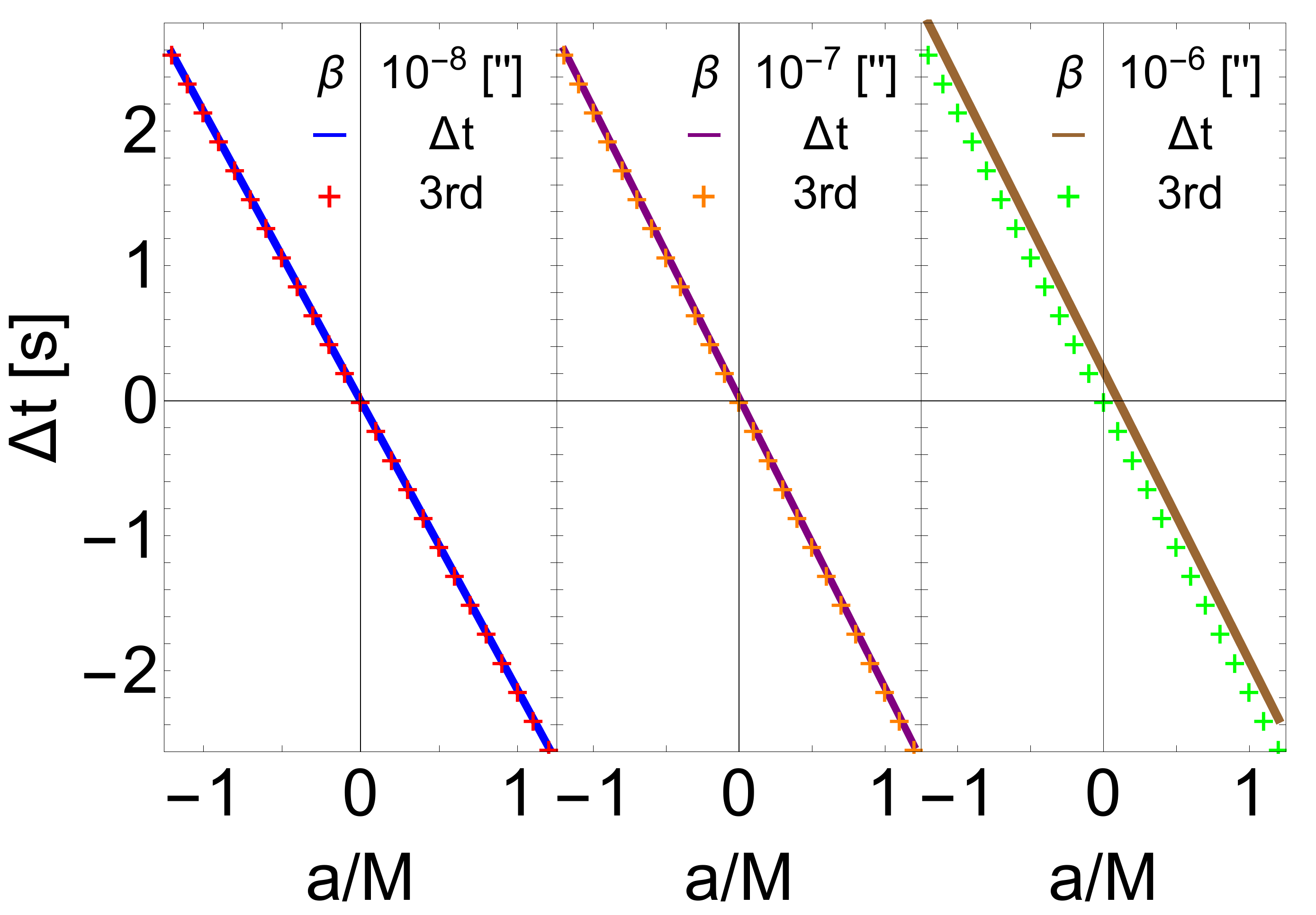}\\
~~~~~(c)\hspace{7cm}(d)
\end{center}
\caption{The time delay \eqref{eq:dtkn} for the KN spacetime as a function of $\beta$ ((a) logarithmically and (b) linearly), $a$ ((c) logarithmically and (d) linearly). The ``1st, 2nd'' and ``3rd'' stand for corresponding terms of Eq. \eqref{eq:dtkn}. In plot (c), the 3rd term depends on $\beta$ very weakly and therefore the solid curves overlap for different $\beta$. In plot (d), the 3rd term dominates $\Delta t$ when $\beta$ is small and therefore the ``+'' lines slightly overlap the solid $\Delta t$ lines. \label{fig:sgra1}}
\end{figure}

In Fig. \ref{fig:sgra1}, we plot the total time delay \eqref{eq:dtkn} and each term of it for the Sgr A*, using values of parameters $M=4.1\times 10^6 M_{\odot}$, $r_{d}=8.12$ [kpc] and $Q \leqslant 3\times 10^8 $ [C] \cite{Andreas:2018}. In Fig. \ref{fig:sgra1} (a) and (b), we plot the time delay logarithmically for $|\beta|<10$ [$^{\prime\prime}$] and linearly for $|\beta|<10^{-5}$ [$^{\prime\prime}$] respectively using value of $a=0.71M$ for a source that is located at radius $r_{S39}$ for light signal $(v=1)$.
This value of $a$ is the mean value of its currently favored range $[0.5M,~0.92M]$ \cite{Andreas:2018}. It is clear that when $|\beta|\lesssim 10^{-3}$ [$^{\prime\prime}$], the third term of Eq. \eqref{eq:dtkn} depends on $\beta$ very weakly. More importantly, when $|\beta|\lesssim 1.5\times 10^{-5}$ [$^{\prime\prime}$], this term is larger than the first and second terms and therefore dominates the total time delay. While for $ 1.5\times 10^{-5}\lesssim |\beta|\lesssim 10^{-3}$ [$^{\prime\prime}$], the first and second terms will be of similar size which is much larger than the third term. When $|\beta|\gtrsim 10^{-3}$ [$^{\prime\prime}$], the first term will be larger than the second and third terms.

To see more clearly the effect of the spin $a$ on the time delay, in Fig. \ref{fig:sgra1} (c) and (d), the time delay are plotted logarithmically and linearly respectively for three values of $\beta$ ($10^{-8}$ [$^{\prime\prime}$], $10^{-7}$ [$^{\prime\prime}$] and $10^{-6}$ [$^{\prime\prime}$]) for $a$ from $-2M$ to $2M$ for light signal. Although it is generally believed that the spacetime in the Galaxy center is a BH rather than a naked singularity, the time delay formula found in this work can work for both BH and naked singularity spacetimes.
It is seen that when $|\beta|<10^{-6}$ [$^{\prime\prime}$], the time delay due to the spin of the spacetime will dominate the total time delay for $|a|\gtrsim 0.2M$. The smaller the $\beta$, the larger the range of $|a|$ in which the spin term dominates. In general, for $|\beta|<10^{-6}$ [$^{\prime\prime}$], it is seen that the time delay depends on the value of $a/M$ linearly, which reaches about 2.0 second at $a/M=1$ for the given parameter setting, as dictated by Eq. \eqref{eq:tdknps}.  This value is well within the reach of current observatories for GRB, GW \cite{TheLIGOScientific:2017qsa,Monitor:2017mdv} or supernova neutrino signals \cite{Jia:2017oar,Suzuki:2019jby}. By measuring this time delay, therefore the spin of the corresponding Sgr A* SMBH can be deduced. We emphasis that this linear dependance of the time delay on spin $a$ for small $\beta$ is not limited to light signals. From Eq. \eqref{eq:tdknps}, we see clearly that the linear dependance is present in the time delay of signals with any fixed velocity, including that of GWs and neutrinos.

Indeed, for relativistic timelike signals such as supernova neutrinos and massive GWs, as can be seen from Eq. \eqref{eq:tdexpv} their time delay will be very close to that of the light signal. The difference of the GW time delay and that of the GRB has been proposed to constrain the GW speed \cite{Fan:2016swi, Liao:2017ioi}.
However, a simple order estimation reveals that this difference, $\Delta t_{c-v}$ in Eq. \eqref{eq:tdexpv}, is much smaller than the main time delay $\Delta t(v=1)$, especially when $v$ is very close to 1 as for supernova neutrinos \cite{Tanabashi:2018oca} and GWs \cite{Abbott:2017oio,TheLIGOScientific:2017qsa,Monitor:2017mdv}.
For KN spacetime, substituting the coefficients \eqref{eq:kncoeff1} into the $\Delta t_{c-v}$ in Eq. \eqref{eq:tdexpv}, one obtains the time delay difference
\be
\Delta t_{c-v}=
\lcb\frac{4 \epsilon M \lsb 3 \eta_K(\beta,1)+ 2 \rsb }{\eta_K(\beta,1) \sqrt{\eta_K(\beta,1)+1}}
-\frac{4\epsilon M }{\sqrt{\eta_K(\beta,1)+1}}
+\frac{2 a \lsb \eta_K(\beta,1)+2 \rsb  \sqrt{M (r_d+r_s)} }{ \sqrt{ \eta_K(\beta,1) \lsb \eta_K(\beta,1)+1\rsb r_d r_s}}\rcb (1-v)
\label{eq:tddkn}
\ee
with $\eta_K(\beta,1)$ given in Eq. \eqref{eq:knbeta} with $v=1$.
Simple order analysis shows that according to the values of $\beta^2r_{s,d}$, $a^2/r_{s,d}$ and $M$, there will be three cases for the dominance of term(s) of Eq. \eqref{eq:tddkn} and the value of $\Delta t_{c-v}$. When (1) $\beta^2r_{s,d}< a^2/r_{s,d}$, the third term of Eq. \eqref{eq:tddkn} dominates and  $\Delta t_{c-v}$ is roughly a constant, $2a\sqrt{M/r_{s,d}}$. If (2) $a^2/r_{s,d}< \beta^2r_{s,d}< M$, the first and second terms contribute similarly and $\Delta t_{c-v}$ will be $2\beta^1(2Mr_{s,d})^{1/2}$. If (3) $\beta^2r_{s,d}> M$, then the first term contribute most to $\Delta t_{c-v}$, which approximately is $\beta^2 r_{s,d}$.
The later two cases are similar to the SSS case in which there is no spin angular momentum \cite{Jia:2019hih}.

\begin{figure}[htp]
\begin{center}
\includegraphics[width=0.6\textwidth]{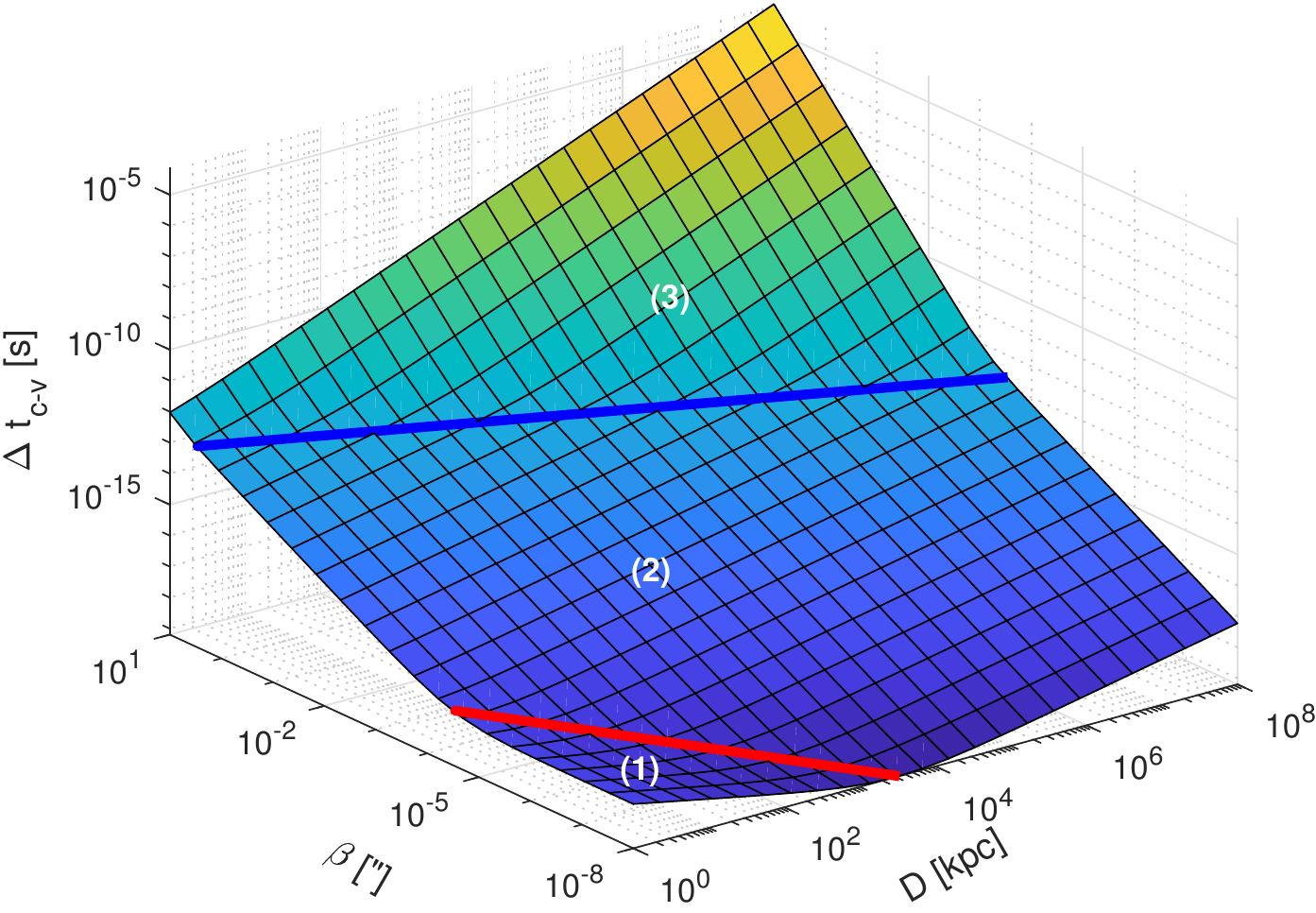}
\end{center}
\caption{The time delay difference \eqref{eq:tddkn} for the KN spacetime as a function of $\beta$ and $r_s=r_d=D$ for a GW with speed $v=(1-3\times 10^{-15})c$. The red and blue curves are plotted using $\beta^2 r_{s,d}=a/r_{s,d}$ and $\beta^2 r_{s,d}=M$. \label{fig:tddiff}}
\end{figure}

In Fig. \ref{fig:tddiff}, we plot $\Delta t_{c-v}$ between time delays of light signal and a timelike signal with velocity $v=(1-3\times 10^{-15})c$ as a function $\beta$ and $r_s=r_d=D$. The $3\times 10^{-15}c$ is the maximal deviation that GW speed can get from the speed of light \cite{TheLIGOScientific:2017qsa, Monitor:2017mdv}. The above three cases of dominance are then separated by the red and blue curves, which are plotted using $\beta^2 r_{s,d}=a/r_{s,d}$ and $\beta^2 r_{s,d}=M$ respectively. As discussed above, in the regions below the red curve (or above the blue curve), the third term (or first term) dominates the time delay difference of Eq. \eqref{eq:tddkn}. While in the region between the two curves, the first and second terms contribute similarly.
Given that the time resolution of current GRB and GW can roughly reach the 0.05 [s] and 0.002 [s] \cite{Monitor:2017mdv} level respectively, from Fig. \ref{fig:tddiff} we observe that in order to obtain $\Delta t_{c-v}$ that is larger than these resolutions, the parameters $\beta$ and $r_{s,d}$ should be in case (3), well beyond the case (1) in which the spacetime spin plays a role. Therefore we can conclude that for the current or near future GRB and GW time resolution, the spin angular momentum will not affect the detectable time delay difference between the two kinds of signals in the weak field limit.

\section{Conclusion and discussions}

It is shown using a perturbative method that the total travel time for both timelike and null signals in the asymptotically flat SAS spacetimes can be expressed as a quasi-series of the impact parameter. The $n$-th order coefficient of this series is determined by coefficients up to the $n$-th order of the asymptotic expansion of the metric functions. Solving the impact parameters for both the GL images, we can obtain the time delay between them to the leading order(s) of $M/r_{s,d}$ and $\beta$ for signals with general velocity. Dominance of different terms in the parameter space spanned by $(M/r_{s,d},~\beta)$ is analyzed. One particular interesting case is when $\beta$ is very small. The time delay in this case then depends on the spacetime spin angular momentum $a$ linearly (see Eq. \eqref{eq:tdknps}). For Sgr A* and typical values of $a$ (e.g. $a=0.71M$), $\Delta t$ can reach the $\mathcal{O}(1)$ [s] order for $\beta\lesssim 10^{-5}$ [$^{\prime\prime}$], well within the precision of time measurement for GW, GRB or neutrinos events. Therefore this time delay might be used to constrain the spacetime spin very precisely. Of course this order of time delay is obtained for source that roughly at the distance of S-stars ($r_{S39}$ to be precise). For this distance, the apparent angles obtained using Eq. \eqref{eq:thetabt} are $\theta_{b}=0.756$ [mas] and $\theta_t=0.756$ [mas] respectively, which are beyond the current resolution of the observatories.
On the other hand, we also emphasis that the time delay itself, that is without having to distinguish the angular position of the two images, will reveal the spin information for us. As can be seen from Eq. (54), in principle we do not have to know the apparent angles of the images in prior in order to deduce spin from the time delay, although being able to resolve them would be most helpful.

Although we mainly applied the time delay to Sgr A*, in principle it can also be applied to other systems, including BHs with regular mass inside the Galaxy, or other SMBHs in nearby galaxy centers. The latter are farther from us than Sgr A*, and we see that the time delay  \eqref{eq:tdknps} will also be larger by a factor of $\sqrt{r_{SMBH}/r_{Sgr~A*}}$. This will ease the time measurement to allow larger uncertainty of the events.

We also emphasis that both the total time \eqref{eq:ttynln} and the time delay \eqref{eq:tdgenformbeta} are applicable to general asymptotically flat SAS spacetimes. Keeping higher orders in $M/b$ and/or $b/r_{s,d}$ than Eqs. \eqref{eq:ttynln} and \eqref{eq:tdgenformbeta}, higher order coefficients in the metric functions will affect the total time or time delay. Therefore by measuring the time delay to the necessary accuracy, these coefficients, which are often of great interest to many alternative gravitational theories \cite{Will:2014kxa}, can be constrained. In addition, we also note that time advancement is another direction to measure quantities about spacetime \cite{Deng:2017umx,Deng:2018ncg}, and it is worth to be studied in the future.

\acknowledgments
We thank Ke Huang for valuable discussions. This work is supported by the NNSF China 11504276.

\end{document}